# APPLICATION OF THE LEARNING FROM ERRORS PRINCIPLE IN TUFTING MACHINES

*Longxiang Shao\*, Dominik Huesener, Michael Schluse, Juergen Rossmann*

*Institute of Man Machine Interaction, RWTH Aachen University, Aachen, Germany*
*\* Shao@mmi.rwth-aachen.de*

**Keywords**: DIGITAL TWIN, MBSE, AR SYSTEM, TUFTING MACHINE

## Abstract

The principle of learning from errors is pedagogically powerful but often impractical in industrial settings due to risks to safety and equipment. This paper presents an integrated training approach specifically designed for tufting machine operators. It uses hybrid digital twins, augmented reality (AR), and Petri Net-based modelling to apply the learning from errors principle effectively. Operator actions and errors are simulated via experimentable digital twins (EDTs), and the consequences of errors are visualized in AR, enabling safe, experiential learning. A Petri Net model formally represents the process, including typical faults and recovery paths, and is implemented in VEROSIM using SOML++. This hybrid framework provides a scalable foundation for AR-guided training systems that reduce risk and accelerate skill acquisition.

## 1  Introduction

It was shown that humans can learn effectively through making errors, realizing their errors, and correcting them. [1], [2] However, in industrial environments, operational errors may result in damage to machines, products, or operators, and thus making errors is generally discouraged. A previous project has shown, for a CNC machine and an injection moulding machine, how learning from errors can be incorporated without the downsides by detecting operator errors and visualizing consequences virtually using AR technology [3]. Section III-A presents the previous project.

Tufting machine operators are often unskilled workers. In order to maximize productivity and minimize errors, adequate operator training is important. Traditional training methods often fail to provide hands-on experience without risking equipment damage or safety hazards. This paper aims to bridge this gap by presenting an innovative integrated training approach based on hybrid digital twins, augmented reality, and Petri Net-based modelling. We hypothesize that a learning system similar to the one described above can significantly shorten the training period for new personnel by allowing the trainee to effectively learn typical setup and operation procedures in a safe environment and using errors as a learning opportunity. A mock-up of the system is shown in Fig. 1.

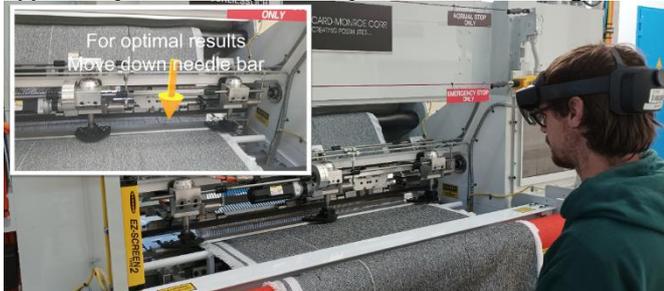

Fig. 1. Photograph showing a mock-up, which provides AR-based instructions and hints to tufting machine users.

## 2.  Theoretical Background

*2.1 MBSE*

Model-Based Systems Engineering (MBSE) provides a structured approach to modelling complex interconnected systems, such as those involving digital twins of operators, machines, and tools, as well as the associated work processes. By leveraging MBSE, the development of the assistance and learning system can be enhanced by systematic determination of the requirements, structure, and behaviour of these systems. MBSE, typically based on standards like UML or SysML, offers a robust methodology for designing, structuring, and formally describing digital twins. This methodology allows for the detailed identification and description of individual digital twins and their components within the tufting machine ecosystem. For instance, SysML blocks within a block definition diagram can be used to specify the behaviour, properties, parameters, dependencies, and functional relationships of these digital twins. The definition of ports and their connections can further specify the data flow between these blocks, ensuring seamless interaction between the digital twins of operator, machine, product, and tools. Additionally, MBSE enables the definition of requirements, which can be visualized using a requirement diagram. This iterative process allows for the step-by-step structural composition of the digital twins, ultimately improving the accuracy and effectiveness of the assistance and learning system. [4], [5]

Modularization, a key aspect of MBSE, significantly enhances the adaptability of the assistance and learning system for tufting machines. By structuring the system into distinct and interchangeable modules, it becomes easier to tailor the system to different work environments or machine models. For instance, when introducing a new tufting machine model, only the relevant modules need to be updated or replaced. This flexibility not only reduces development time and costs but also ensures that the system remains up-to-date and relevant across diverse operational settings.

## 2.2 EDT

The Experimentable Digital Twin (EDT) merges Digital Twins with advanced simulation capabilities, serving as a powerful framework for both individual and system-of-systems simulations. EDTs are modelled based on the physical architecture of a system, with a focus on its structural composition, which enables comprehensive simulations on a system level while maintaining detailed component-level insights. [5]

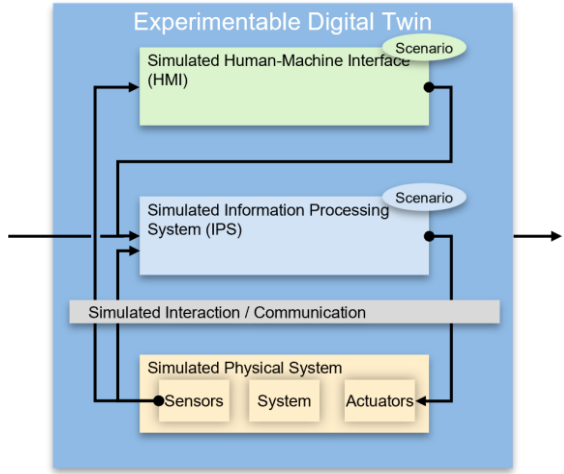

Fig. 2. Basic structure of an EDT [5].

EDTs mirror not just the physical asset but the entire cyber-physical system, including its interactions with the environment and communication capabilities. They are thus suited to analyse, optimize, verify, and validate complex interconnected systems through detailed simulations.

EDTs consist of several components as depicted in Fig. 2: Simulated Physical Asset (SPA), Simulated Data Processing System (DPS), and Simulated Human-Machine Interface (HMI). These components communicate via a simulated communication infrastructure, which can be connected to the real system if necessary. The SPA is the digital representation of the system and includes sensors (e.g. cameras) and actuators (e.g. motors). The simulated DPS processes the data from the virtual sensors and performs actions through the simulated actuators. Both SPA and DPS are controlled and monitored through the simulated HMI. It provides functionality comparable to the real HMI to manipulate the physical system virtually.

EDTs can model complex application scenarios involving multiple interacting EDTs at different levels of detail. These scenarios provide insights into the overall system behavior, which might not be apparent from simulating individual subsystems.

Experimentable Digital Twins (EDTs) are particularly valuable in AR-based learning systems due to their ability to simulate real-world scenarios in a controlled virtual environment.

- **Safe Learning Environment:**

EDTs allow trainees to interact with virtual replicas of complex systems without the risk of damaging expensive equipment or causing damage. This is crucial for learning scenarios involving dangerous or costly operations.

- **Realistic Simulations:**

By replicating the behaviour and interactions of real systems, the EDTs provide realistic simulations that can be used to demonstrate the consequences of user actions in AR. This enhances the learning experience by making it more immersive and practical.

- **Immediate Feedback:**

EDTs can simulate the results of user actions in real time, providing immediate feedback. This is essential for AR-based learning, where users need to instantly see the results of their interactions to understand the impact of their decisions.

- **Integration with AR Technology:**

The detailed models provided by EDT can be integrated with AR technology to overlay virtual information onto the real world, enhancing the learning experience by providing contextual and interactive content.

In a previous project, a digital twin of a tufting machine was created that used a kinematic model of the machine constructed from technical drawings, which includes a user interface which allows the user to set all the parameters within the bounds of the real machine. The digital twin allows the user to study the interaction of gripper, needle, and knife in detail, helps the user to find optimal settings and transfer them to the real twin.

In summary, EDTs enhance AR-based learning systems by providing a safe, realistic, and interactive environment for trainees to practice and learn, with immediate feedback and customizable scenarios that can be tailored to specific training needs.

## 3 AR-based Learning Systems

Yin et al. [6] conducted a literature review on AR-assisted digital twins and found 9 papers that focused on using AR-based DTs for operator training in transportation and engineering education, including case studies on electric circuits, excavators, robots, a production line, laboratory multi-tank system and modular construction system, Internet of Things (IoT) device visualization and control, as well as turbine design. The AR system is useful to simulate behaviour outcomes based on real data, but with increased security.

David et al. [7] envisioned a framework that uses Digital Twins to provide hands-on training in production-based systems. They highlight the role that computer simulations and virtual reality have played in manufacturing education. Their didactic methodology incorporates Digital Twins for university level production engineering education that are demonstrated during lectures. In laboratory exercises the student can experi-ment with the digital twin, and on the physical site, a physical demonstration is possible. Their didactic framework is based upon learning theories such as Cognitivism, which states that learning requires active participation and actions performed based on cognition.

### 3.1 FeDiNAR

The FeDiNAR [3] project (German acronym for "failure-driven industrial training with augmented reality") redefines operational errors as valuable pedagogical assets, developing an AR-based learning system with contextualized training scenarios to maximize error-driven skill acquisition. Adopting a human-centric pedagogical approach over reliability-oriented engineering methods, this study focuses exclusively on human-induced errors that originate from workflow actions and propagate through technical systems. To align with this paradigm, stochastic technical components (e.g., faults [8] and failures [9]) are excluded through deterministic system modelling.

Since the FeDiNAR system can only detect discrete action states, additional information is required to perform a causal error analysis. Inspired by Kapur's learning concepts [10], the project incorporates a debriefing or feedback phase following apprentices' interaction with the system. This debriefing stage remains essential even when the apprentices provide correct solutions. To facilitate this, FeDiNAR defines system-observable events and behaviours, discretizing continuous workflows. The dynamic behaviour of the FeDiNAR system is formally modelled using Time-extended Petri nets. When a learner performs an action (corresponding to a state in the Petri net) and fails to achieve predefined objectives, the system assigns a new state based on historical operations. This unmet goal state is defined as an error.

For data collection, FeDiNAR employs a HoloLens (AR head-mounted display) to capture user operations and sensors on primary or auxiliary physical equipment to gather system status information. For error consequence simulation, dual modalities are employed:

- **White-box simulation:**

Physics-based simulation engines generate authentic system responses.

- **Black -box simulation:**

Pre-programmed scenarios designed with specific educational objectives, not necessarily reflecting actual physical phenomena.

This integrated framework establishes a closed-loop learning pathway encompassing error detection, causal analysis, and corrective training.

*3.2 FischerTwin*

FischerTwin [4] is the first demonstrative digital twin system representing the core concepts of the FeDiNAR project, based on a modular Fischertechnik swivel-arm robot. A framework integrating the EDT and the Virtual Testbed (VTB) is proposed and its design methodology and implementation process are described in detail. Specifically, guided by the ECSS project phase definition and MBSE, the research team has carried out function definition, requirements analysis, structural system design, and developed high-fidelity digital twin submodels, thereby establishing a comprehensive joint simulation system combining EDT and VTB.

FischerTwin enables high-fidelity virtual testing across multidisciplinary domains, including kinematics, rigid body dynamics, sensor signals, communication interfaces, and electrical wiring, significantly reducing reliance on physical prototype development and associated costs. Utilizing this framework, the research team concurrently developed a robotic command interpreter, an OPC UA communication interface, and a graphical human-machine interface, all of which were seamlessly transferred to the physical system at minimal cost. Furthermore, FischerTwin supports flexible switching between real-time mirroring and predictive simulation modes.

It can synchronize with the real system state in real-time, perform temporal backtracking, and conduct predictive simulations, enabling effective virtual rehearsals, assessments, and validations.

Overall, FischerTwin demonstrates the feasibility and significant potential of FeDiNAR's EDT and VTB cooperative framework throughout the entire life cycle of complex cyber-physical systems, particularly highlighting its value in safety assessment, rapid iterative development, and predictive state analysis.

## 4 Concept/Framework

*4.1 Requirement*

When applying the core principles of the FeDiNAR project to a tufting machine, the system must meet a set of critical functional and interaction-related requirements. First, the system shall provide a comprehensive and accurate digital twin representation that encompasses the physical structure of the machine, operational processes, run-time conditions, potential failures, and their consequences.

Furthermore, the system shall be self-explanatory and integrate seamlessly into the working environment without requiring additional training. To support intuitive understanding, a simple and user-friendly visualization approach should be provided to help users quickly grasp the structure and behaviour of the system.

The system should also enable timely error detection through suitable sensors and offer real-time feedback in response to potentially hazardous actions. This includes the capability for manual intervention, such as automatically stopping the machine and displaying the resulting consequences.

In addition, the system shall implement a logging mechanism to record all user interactions, enabling retrospective analysis and reflection. Together, these requirements ensure that the digital twin delivers educational value, operational safety, and high usability in real-world industrial applications.

The requirements are structured in an MBSE fashion and visualized in the requirements specification diagram shown in Fig. 3.

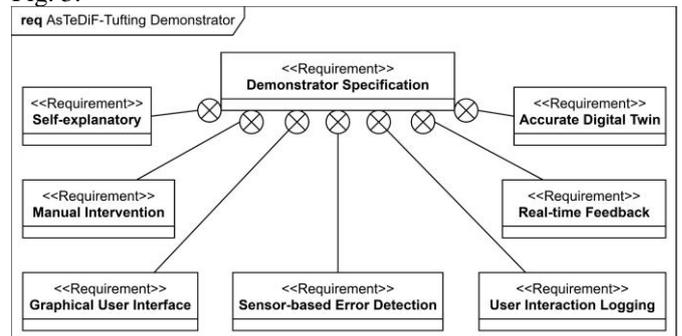

Fig. 3. SysML requirements specification diagram for the AsTeDiF-Tufting demonstrator.

*4.2 Hybrid Twins of the Working Environment*

Hybrid Twins combine physical and virtual components to create a comprehensive digital representation of the working environment. This integration allows for real-time monitoring, error detection, and assistance during the tufting process.

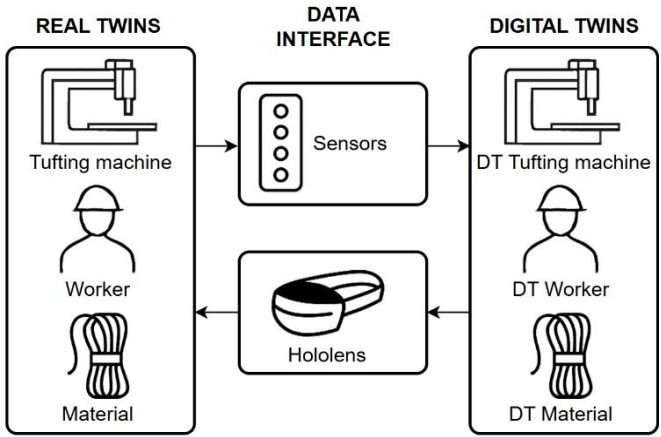

Fig. 4. Hybrid twins framework for the AsTeDiF-Tufting demonstrator. Sensors monitor the status of material, tufting machine and worker, and update the according digital twins. Deviations from the expected state can be visualized through the HoloLens.

The working environment encompasses several elements, each with its physical and virtual counterparts. The elements can be divided into the work objects, work person, and work equipment. The work person uses work equipment as an auxiliary means to act on work objects.

At the core is the tufting machine itself, which includes the tools: needles, grippers, and knives. The virtual component of the tufting machine is represented by a detailed CAD model that includes parameters like turning speed, pile height, and machine status (off, run-mode, etc.). Sensor values for needles, grippers, and knives are continuously monitored, along with the main shaft rotational velocity, to ensure accurate and up-to-date information.

Another crucial element is the creel with spools, which holds the yarn used in the tufting process. The virtual twin of the creel keeps track of whether each spool position is empty or occupied, the type of yarn on each spool, and whether it is connected to the tufting machine or blocked by a tension thread. Wrong spools or tension threads are visible on the tufting product, which either shows a regular pattern if everything is in order or an interrupted pattern if some error occurred. The substrate, onto which the tufting is applied, is also a key component. The virtual twin of the substrate includes data on the material type, length, and the position of any seams.

The operator plays a crucial role in the working environment, and their interactions with the machine are closely monitored. Eye-tracking data determines the operator's focus, while hand-tracking data monitors the operations performed.

This information is used to determine operation errors and interfere e.g. when an error is unnoticed and can cause damage to product and/or machine.

Additional tools and materials, such as compressed air and yarn splicing equipment, are integrated into the Hybrid Twin system where necessary. For example, the duration of applying compressed air can be recorded and analysed to ensure it meets the requirements.

Data collection is a critical aspect of maintaining an accurate digital twin. Cameras and image recognition can be used to monitor the physical components and the operator's actions. Encoders can provide precise data on the position and movement of machine parts, while light barriers can detect the presence or absence of objects. The machine status memory records the operational state of the tufting machine, ensuring that all relevant data is captured.

Changes detected through sensors are mirrored to the EDT in VEROSIM, a 3D rigid body simulation system developed at the Institute for Man-Machine Interaction. An accurate model of the tufting machine environment and the simulation capabilities of VEROSIM allows predictions based on the current state and recognized intended user actions. The different Hybrid Twins can interact in the VEROSIM VTB.

With Hybrid Twins, the working environment of tufting machines can be closely monitored, thus enabling error detection. This is the basis for the desired learning system for tufting machines.

### 4.3 Partitioned Activity Petri Net Architecture

In the context of learning from errors training for tufting machines, it is essential to model and monitor operator actions in real time so that feedback or safety measures can be triggered as soon as an error is about to occur. Petri Nets provide a formal modelling framework that is well suited for this purpose, as they naturally capture concurrency, synchronization, and conditional events. A Petri Net consists of places (system states), transitions (state changes or events), and arcs (causal relations). Tokens represent the current state of the system, and their flow illustrates the sequence and concurrency of actions within complex tasks. In this work, we adopt and extend the Partitioned Activity Petri Net Architecture introduced by Herrmann et al. [11] to support real-time recognition of human activities and error feedback in tufting machine operations.

The proposed architecture consists of three subnets: the Abstract Activity Net, the Activity Execution Net, and the Error Consequence Net. Together, these subnets represent both the abstract lifecycle of human activities and the concrete execution of actions, while also enabling the system to respond appropriately when errors occur. The Abstract Activity Net (Fig. 5) models the abstract lifecycle of an activity, focusing on key phases such as Start – Execute – Finish – Interrupt. Activities to be executed are first stored in the Pool, representing pending activity instances. An activity can only begin when its associated Guard condition is satisfied, at which point tokens move from the Pool into the Start transition. The Abstract Activity Net does not interact directly with the sensor data. Instead, it receives subevent signals from the execution level through fusion places, shown as gray ovals in the figure.

Fusion places synchronize tokens between subnets, ensuring that activity events detected in the Activity Execution Net are consistently reflected at the abstract level. When an error occurs, the activity is interrupted, and related information is recorded in the History and Log places. The primary role of the Abstract Activity Net is to record activity states and to provide a unified interface for error detection and assistive actions.

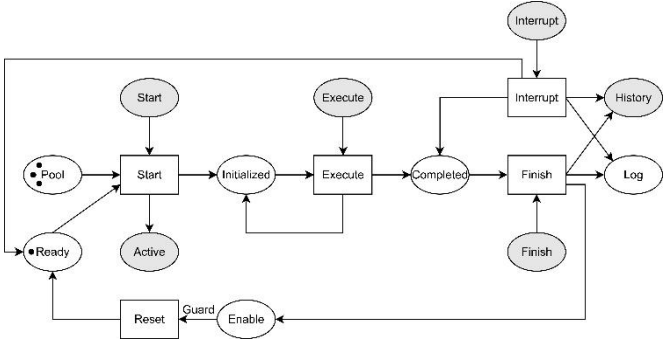

Fig. 5. Abstract Activity Net

The Activity Execution Net (Fig. 6) models concrete actions, such as yarn splicing or mounting new spools. This subnet is driven by sensors and interface functions, which detect subevents such as start, execute, and end. These events are then mapped into tokens that can be used at the abstract level. A time-step counter is used to record the duration of actions. When an error is detected, e.g. a broken yarn, the current operation is interrupted, and the error is written in the History place of the Abstract Activity Net. In this way, the execution net ensures that real-world actions are consistently transformed into abstract-level events, enabling unified tracking of both normal and erroneous outcomes.

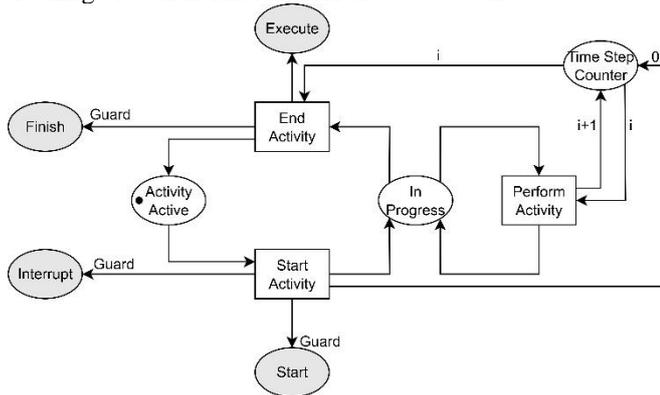

Fig. 6. Activity Execution Net

The Error Consequence Net (Fig. 7) is a dedicated safety subnet within the execution level. When the Abstract Activity Net indicates through the History and Active flags that an activity has been executed and has resulted in an error, this subnet is activated and triggers predefined error-handling measures, such as using augmented reality visualization to present the error to the user or engaging machine interlocks to prevent hazardous consequences. Different tasks can define tailored Error Consequence Nets, ensuring that the overall architecture remains both general and adaptable.

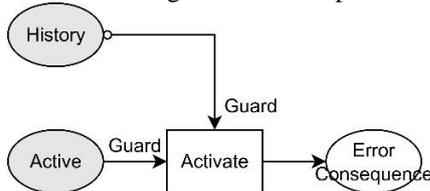

Figure 7. Error Consequence Net

The Petri Net model used in this work rigorously formalizes the training process of tufting machine operators by explicitly specifying error detection and recovery pathways. Implemented in the VEROSIM simulation software in combination with the SOML++ scripting language, it supports simulation-based verification and formal analysis, such as reachability and liveness analysis, providing a solid foundation for AR-assisted tufting machine training systems. At the same time, this modelling framework offers high extensibility, allowing the integration of additional features such as automated yarn tension checks, and visual verification of winding patterns significantly enhancing the practical utility and instructional effectiveness of the training system.

*4.4 TwinCAT, HoloLens, Sensors*

In a previous project, the tufting machine was equipped with sensors that continuously measure the positions of the tools (needles, grippers, and knifes). The sensors use EtherCAT to transfer their data to a laptop that runs TwinCAT. In TwinCAT, sensor values are read and processed. The digital twin in VEROSIM is notified through the TwinCAT Automation Device Specification (ADS) interface whenever sensor values change and adjusts its representation accordingly in real-time. Thus, in the mirroring mode, the digital twin is able to show current tool positions. A visualisation of worksteps and error consequences is shown using a HoloLens. The HoloLens runs a client of the digital twin software VEROSIM which is connected via Wi-Fi to the digital twin. AR allows the trainee to show the consequences of the errors he makes virtually at the correct position where they would occur if the process was not interrupted by the learning system. The AR headset further has the advantage that the trainee still has both hands available for interacting with the real machine and eye tracking can be used to infer what element the user is focusing on.

Some training tasks are also possible without access to the real machine. In this case, static machine parts can be replaced, e.g. by paper mock-ups and variable parts rendered in the HoloLens while the user might still interact with physical tools or a user interface mock-up through a touchscreen. In addition, the machine status (whether it is running, parameters...) is read from the machine itself, which also uses EtherCAT; and additional sensors, such as cameras, when combined with data or image processing algorithms, are used to infer the position and status of further work elements, products, or tools. An example could be checking whether a spool was positioned correctly in a creel using marker detection. A focus of the assistance system is on machine setup.

## 5 Conclusion and Future Work

In conclusion, this paper has presented an innovative approach to operator training for tufting machines by integrating the principles of learning from errors with advanced technologies such as hybrid digital twins and augmented reality. By creating a safe environment for trainees to learn tufting machine operation, it is possible to enhance skill acquisition while mitigating risks associated with traditional training methodologies. The proposed framework not only facilitates experiential learning but also allows for immediate feedback on operator actions, ensuring that mistakes become valuable learning opportunities rather than setbacks.

Currently, initial work was carried out in the development of the Petri Nets as well as object / task recognition. The existing digital twin needs to be altered and DTs of components other than the tufting machine need to be implemented. In addition, visual representations of the error consequences need to be developed. The effectiveness and acceptance of the resulting system is to be tested taking into account the reduction in training time, error rates, skill retention and user satisfaction. Ideally, the system is evaluated through end-user testing and compared to traditional training methods.

Moving forward, based on a detailed analysis of the work, the proposed system will be implemented. A demonstrator is created to evaluate its effectiveness.

## 6 Acknowledgements

The IGF project "Development of an assistance system for tufting machines with didactic use of errors for operational training and learning processes as well as retooling, maintenance and repair processes using the example of the tufting machine (AsTeDiF)" 23479 N of the Research Association Forschungskuratorium Textil e.V., Reinhardtstraße 14-16, 10117 Berlin was funded by the German Federal Ministry of Economics and Climate Protection via the AiF as part of the program for the promotion of joint industrial research and development (IGF) based on a resolution of the German Bundestag.